\documentclass[prl,twocolumn,superscriptaddress,preprintnumbers,a4paper]{revtex4} 
\usepackage{graphicx} 
\usepackage{amsmath} 
\usepackage{amssymb} 
\usepackage{amstext} 
\usepackage{amsxtra} 
\usepackage{version} 
\usepackage{longtable} 
\begin{document} 
%
\newcommand{\Tc}{T$_{\mathrm C}$} 
\newcommand{\Tco }{T$_{\mathrm CO}$\space} 
\newcommand{\sinth}{\mbox{$\sin\theta/\lambda$}} 
\newcommand{\inA}{\mbox{\AA$^{-1}$}} 
\newcommand{\mub}{\mbox{$\mu_{B}$}} 
\newcommand{\mns}{$-$} 
\newcommand{\ddd}{3\textit{d}} 
\newcommand{\pp}{2\textit{p}} 
\newlength{\minusspace} 
\settowidth{\minusspace}{$-$} 
\newcommand{\msp}{\hspace*{\minusspace}} 
\newlength{\zerospace} 
\settowidth{\zerospace}{$0$} 
\newcommand{\zsp}{\hspace*{\zerospace}} 
\newcommand{\lasrmn}{La$_{2-2x}$Sr$_{1+2x}$Mn$_{2}$O$_7$\space} 
\newcommand{\mnt}{$Mn^{3+}$} 
\newcommand{\mnf}{$Mn^{4+}$} 
\newcommand{\eg}{$\textit{e}_{g}$ } 
\newcommand{\qce}{$\textit{q}_{CE}=(1/4,1/4,0)$} 
\newcommand{\qcel}{$\textit{q}_{CE}=(1/4,-1/4,0)$} 
\newcommand{\qos}{$\textit{q}_{L}=(0.3,0,1)$} 
\newcommand{\sqw}{S(\textit{Q},$\omega$)} 
\newcommand{\Ts}{T$^{*}$\space} 
\newcommand{\GG}{$\Gamma$\space} 
\newcommand{\NaH}{Na$_{x}$CoO$_{2}\cdot y$H$_{2}$O} 
\newcommand{\NaD}{Na$_{x}$CoO$_{2}\cdot y$D$_{2}$O} 
\newcommand{\NaCo}{Na$_{x}$CoO$_{2}$} 
\newcommand{\wat}{H$_{2}$O} 
\newcommand{\deut}{D$_{2}$O} 
\newcommand{\etal}{\textit{et al.}\space} 
\newcommand{\degc}{$^{\circ}$C} 
\newcommand{\degg}{$^{\circ}$} 
\newcommand{\br}{Br$_{2}$\space} 
\newcommand{\xrange}{$0.28<x<0.37$} 
\newcommand{\dhx}{$2a \times 2a \times c$} 

\preprint{} 
\date{\today} 
\title{ 
The structure of intercalated water in superconducting 
Na$_{0.35}$CoO$_{2}\cdot$1.37D$_{2}$O: Implications for the superconducting 
phase diagram 
} 

\author{D. N. Argyriou} 
\email[Email of corresponding author: ]{argyriou@hmi.de} 
\affiliation{Hahn-Meitner-Institut, Glienicker Str. 100, Berlin D-14109, 
Germany} 
\author{C. J. Milne} 
\affiliation{Hahn-Meitner-Institut, Glienicker Str. 100, Berlin D-14109, 
Germany} 
\author{N. Aliouane} 
\affiliation{Hahn-Meitner-Institut, Glienicker Str. 100, Berlin D-14109, 
Germany} 
\author{P.G. Radaelli} 
\affiliation{ISIS Facility, Rutherford Appleton Laboratory-CCLRC, Chilton, 
Didcot, Oxfordshire, OX11 0QX, United Kingdom} 
\author{L.C. Chapon} 
\affiliation{ISIS Facility, Rutherford Appleton Laboratory-CCLRC, Chilton, 
Didcot, Oxfordshire, OX11 0QX, United Kingdom} 
\author{A. Chemseddine} 
\affiliation{Hahn-Meitner-Institut, Glienicker Str. 100, Berlin D-14109, 
Germany} 
\author{J. Veira} 
\affiliation{Hahn-Meitner-Institut, Glienicker Str. 100, Berlin D-14109, 
Germany} 
\author{S. Cox} 
\affiliation{Department of Materials Science and Metallurgy, University of Cambridge, Pembroke Street, Cambridge, CB2 3QZ, U.K.} 
\author{N. D. Mathur} 
\affiliation{Department of Materials Science and Metallurgy, University of Cambridge, Pembroke Street, Cambridge, CB2 3QZ, U.K.} 
\author{P. A. Midgley} 
\affiliation{Department of Materials Science and Metallurgy, University of Cambridge, Pembroke Street, Cambridge, CB2 3QZ, U.K.} 

\begin{abstract} 
We have used electron and neutron powder diffraction to elucidate the structural properties of superconducting \NaD. Our measurements show that our superconducting sample exhbits a number of supercells ranging from $\frac{1}{3}a^{*}$ to $\frac{1}{15}a^{*}$, but the most predominant one, observed also in the neutron data, is a double hexagonal cell with dimensions \dhx. Rietveld analysis reveals that \deut\space is inserted between CoO$_{2}$ sheets as to form a layered network of NaO$_{6}$ triangular prisms. Our model removes the need to invoke a 5K superconducting point compound and suggests that a solid solution of Na is possible within a constant amount of water $y$.
\end{abstract} 
\maketitle 

The tuning of Na content in the alkali layered cobaltate Na$_{x}$CoO$_{2}$ results in remarkable changes in its physical behavior, ranging from  magneto-thermoelectricity at $x$=0.75\cite{Wang} to charge ordering at  $x$=0.5\cite{Huang, Muk} and 5K superconductivity at $x$=0.3 after intercalation of \wat.\cite{Takada} This interesting combination of physical 
phenomena occurs within a structural motif reminiscent of geometrically frustrated systems. Here, CoO$_{2}$ sheets are constructed by edge sharing  CoO$_{6}$ octahedra, forming a quasi-2D triangular net. The exact mechanism  of superconductivity within this triangular motif is currently the topic of  extensive scientific inquiry, as it holds the possibility of realizing  Anderson's Resonating Valence Bond model.\cite{baskaran} 

Water insertion increases the separation between CoO$_{2}$ sheets and band structure calculations suggest that this effectively minimizes the electronic inter-planar coupling.\cite{Johannes, Marianetti}This dimensional cross-over effect is corroborated by measurements showing that \Tc\space increases with the separation between CoO$_{2}$ sheets.\cite{Milne,Chen}  However, the same measurements also show that \Tc\space does not vary significantly with electronic doping $x$.  This contradicts a proposed analogy between this alkali hydrate 
and the high-\Tc\space cuprates \cite{Schaak}, whereby optimal \Tc\space would occur within a narrow range of electronic doping.

The resolution of this controversy may lie in the details of the crystal 
structure of this alkali hydrate.  Two mutually incompatible structural models of water-intercalated of \NaH\space, have been proposed to date.\cite{Lynn,Jorgensen} The model proposed by Lynn \etal asserts that the intercalated water assumes a structure similar to \textit{ice}\cite{Lynn} in between the CoO$_{2}$ sheets, whereas the one by Jorgensen \etal suggests a coordination and crystal chemical linkage between Na and \wat\space similar that found in many alkali hydrate systems.\cite{Jorgensen} Both structural models imply that Na concentration $x$ and the amount of intercalated water $y$ are intimately related. But Jorgensen's model also implies that \emph{optimal} \Tc\space corresponds to a line compound exhibiting full Na and \deut\space ordering with a Na/\wat\space ratio $\sim$1/4.\cite{Jorgensen}. Both models are based on the parent structure and unit cell, and rely heavily on chemical constraints to interpret the multiple symmetry-equivalent water sites in that cell.  In spite of their complexity, these models do not reproduce some important aspects of the data.  In particular, an intense Bragg peak at (2.8\AA), and a broad distribution of intensity around ($\sim$2.6\AA) are not accounted for, although Jorgensen \etal speculate that both features may result from a larger periodicity.

To unravel the complexity of its crystal structure and gain insight into parameters that are critical in establishing a superconducting phase diagram, we clearly need to establish the true translational symmetry of \NaH\space.  To this end, we have performed electron diffraction (ED) and neutron powder diffraction (NPD) measurements, the latter  employing isotopic H-D substitutions. Both sets of data clearly establish that the \emph{dominant} ordering mode is a \emph{doubling} of the hexagonal unit cell along both $a$ and $b$.  Our analysis shows that Na-atoms coordinate with \deut\space to form an ordered network of NaO$_{6}$ triangular prisms. This model can account well for the intense Bragg reflection in the neutron data at $\sim$2.8\AA. ED also shows a number of different superstructures, suggesting significant Na-inhomogeneities. These additional ordering modes are most likely the origin the of a broad distribution of scattered intensity at $\sim$2.6 \AA, a common feature of all the neutron data published thus far.  Our double-cell model is able to accommodate a variable amount of Na within the same water framework, strongly suggesting that \NaH\space is \emph{not} a line compound.  Our evidence of more than one Na/water ordering mode within a presumably coherent CoO$_{2}$ network should open a serious debate about the true nature of the superconducting phase.

Description of the synthesis and characterization of these samples is 
found elsewhere.\cite{Milne} In this work two samples were investigated with $x$=0.35, which exhibited a superconducting transitions at 4.5K and 2.8K.\cite{Milne} ED patterns of these samples were taken using a Philips CM30 transmission electron microscope operated at 300kV with the sample cooled to approximately 90K using a Gatan double tilt liquid nitrogen stage. NPD data were measured from these polycrystalline samples at 2K using the high resolution powder diffractometer E9 ($\lambda$=1.7973\AA), located at 
the Berlin Neutron Scattering Center, at the Hahn-Meitner-Institut. Additional data were collected on samples  of varying $x$ on the GEM diffractometer at ISIS which were critical in developing the present model, and will be presented elsewhere. To located the O-atoms of the water molecules,  the sample with \Tc=2.8K that was originally hydrated with \deut, was exposed to a (\wat)$_{0.641}$ (\deut)$_{0.359}$\space moisture mixture at a temperature of 26\degc\space for 3 days. This isotopic H/D ratio gives a zero mean coherent neutron scattering length for protons ($\overline b_{H}$=-0.374,$\overline b_{D}$=0.667), so in principle for this sample the contribution of protons to the entire diffraction pattern can be ignored.  

\begin{figure}[!t] 
\includegraphics[natheight=725,natwidth=600,trim= 80 150 0 10,scale=0.6]{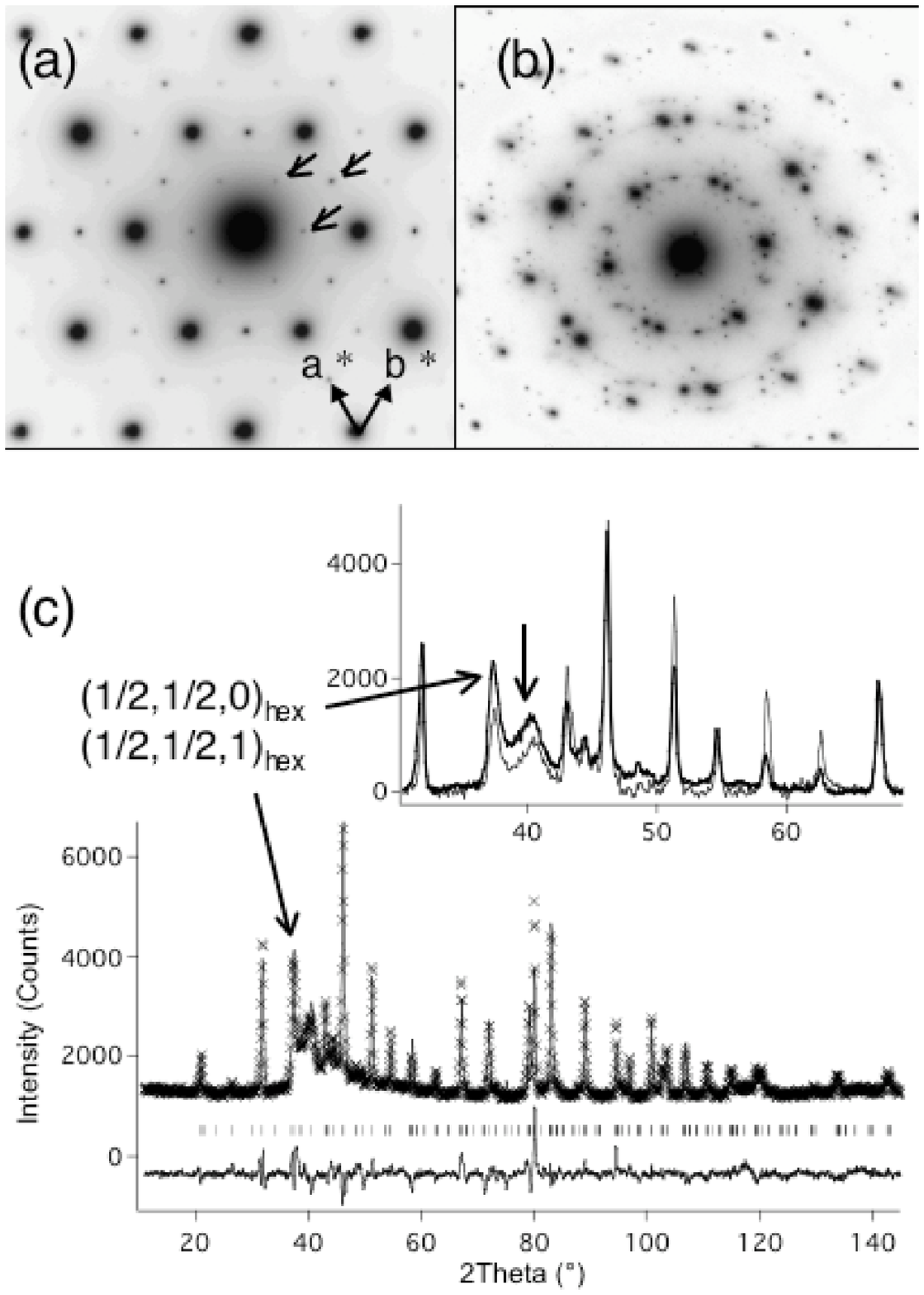} 
\caption{(a-b) ED patterns from Na$_{0.35}$CoO$_{2}$.y\deut. The arrows in 
(a) indicate the cell doubling reflections.  (c) Neutron powder diffraction 
data measured at 2K from  Na$_{0.35}$CoO$_{2}\cdot$yD$_{2}$O. The difference between the observations and the model are plotted under the diffraction pattern, while the expected location of reflection are shown as vertical lines in between. To account for the broad scattering at centered at 40\degg\space (indicated by arrow in inset) the background was interpolated manually for the entire pattern. The inset in (c) depicts a portion of the neutron diffractogram  from the Na$_{0.35}$CoO$_{2}\cdot y$\deut\space (heavy line) and Na$_{0.35}$CoO$_{2}\cdot y$(H$_{0.641}$D$_{0.359}$)$_{2}$O (light line). The background arising from the incoherent neutron scattering from H-atoms has been subtracted. 
} 
\label{diffraction1} 
\end{figure}

Our ED measurements show that the \emph{average} symmetry of the superconducting compound remains hexagonal everywhere.  In the majority of grains, super-lattice reflections indicate a cell doubling giving an average unit cell of \dhx\space (see fig.~\ref{diffraction1}a) where $a$ and $c$ are the lattice constants of the parent P$6_{3}/mmc$ hexagonal cell. Examination of the NPD data shows the strong reflection observed at 37\degg (2.8\AA) can be assigned to the ($\frac{1}{2},\frac{1}{2},0$), ($\frac{1}{2},\frac{1}{2},1$) Bragg peaks, as shown in fig.~\ref{diffraction1}c.  We can construct a chemically sensible double-cell structural model that fits the 2.8\AA\space peak, providing the strongest indication that this and the ED superlattice reflections have the same origin. Further investigation of the ED data reveals an even more complex superstructure in some areas of the sample as seen in fig. ~\ref{diffraction1}b.  Satellite reflections, decorating the parent hexagonal lattice reflections, are arranged in hexagonal nets with q-vectors whose magnitude has been seen to vary from 1/15$\mathbf{a^{*}}$ to 1/3$\mathbf{a^{*}}$, but whose direction is always parallel or nearly parallel to $<110>^{*}$.  The example shown in fig ~\ref{diffraction1}b corresponds to a supercell translation vector of 3$a$ + 3$\frac{2}{3}b$, requiring a \emph{discommensuration} to allow registry between the supercells, equivalent to the Na1-Na2 interatomic vector (see below).  In all such patterns mirror symmetry perpendicular to the basal planes is broken as seen by the gross asymmetry in the intensities of satellite reflections of type $\mathbf{g\pm q}$.  However, the 6-fold symmetry appears to be conserved throughout.  The detailed superstructure of these modulated phases will be discussed further in a forthcoming publication. 

These more complex super-structures seen in the ED 
measurements are not clearly observed in our NPD data  
Indeed, an examination of the NPD data does not show the presence of any additional super-lattice reflections above 37\degg\space suggesting that disorder (which we model as a large effective Debye-Waller factor for water) is substantially attenuating their intensity (superstructure 
reflections in ED are enhanced due to dynamical diffraction).  Comparison of the diffractograms from our deuterated and isotopically mixed HD samples show appreciable differences as illustrated in the inset of fig. ~\ref{diffraction1}c. While some reflections show no significant change in their intensity with the isotopic substitution (for example 2$\theta\sim$31 and 37\degg), others do. That the intensity of the super-lattice reflections at 37\degg\space as well as the broad feature at 40\degg\space is changed, indicates that their structure factor has an appreciable contribution from protons. The origin of the broad diffuse feature at 40\degg\space (2.6\AA) is controversial  as it has been described either as an impurity phase\cite{Lynn} or a broad Bragg reflection\cite{Jorgensen}.  We argue below that it may arise from a superposition of Bragg peaks from the additional ordering modes and from short-range order.

To construct a structural model of the \NaD\space superconductor we analyzed the NPD data using a double hexagonal unit cell (\dhx). The space group was identified as P$6_{3}/m$ using symmetry analysis  of the parent P$6_{3}/mmc$ space group and taking into consideration the ED observations. The solution of the crystal structure was achieved by first analyzing  the isotopically exchanged H-D sample to located the positions of the O-atoms of the water molecules, and then the deuterated sample to located the positions of the protons. This approach produced a good agreement between our model and the NPD data, including the intensity of the reflections at 37\degg as shown in fig.~\ref{diffraction1}c. Structural parameters and occupancies determined from the Rietveld analysis of the deuterated sample are given,  in table ~\ref{NPDresults}. 

\begingroup 
\begin{table}[!bt] 
\squeezetable 
\caption{\footnotesize Structural parameters determined from Rietveld 
analysis of our NPD data measured from our deuterated  $x=0.35$ sample at 2K. The refined lattice constants were $a=$5.63477(21)\AA\space and $c=$19.5456(13)\AA\space (space group P$6_{3}/m$). The weighted profile R-factor for this refinement is $wRp$=6.22\%. The occupancies of the Na sites were constrained so that $f(Na1)+f(Na2)=0.35$. The deuterons labeled D31 and D32 are coordinated to O3, and similarly for O4,D41,D42. The refined total amount of \deut\space was computed to be $y$=1.37(1). In the Rietveld analysis \deut\space was inserted as a rigid body, with a D-O bond-length constrained to 0.99 \AA\space and the D-O-D torsion angle set to 108\degg. $U_{iso}$(Co,O,Na)=0.0005(2)\AA$^{2}$ and $U_{iso}$(\deut)=0.035(4)\AA$^{2}$. Selected bondlengths (in \AA) are given at the bottom of the table. } 
\label{NPDresults} 
\begin{center} 
\begin{tabular}{lllll} 
\footnotesize 
Atom    &    $x$    &    $y$    &    $z$    &    $f$    \\ 
\hline 
Co1    &    0    &    0    &    0    &    1.0    \\ 
Co2    &    $\frac{1}{2}$    &    0    &    0    &    1.0    \\ 
O1    &    $\frac{1}{3}$    &    $\frac{2}{3}$    &    -0.0463(5)    & 
1.0    \\ 
O2    &    0.1733(13)    &    0.3329(15)    &    0.04756(22)    &    1.0 
\\ 
Na1    &    0    &    0    &    $\frac{1}{4}$    &    0.84(4)    \\ 
Na2    &    $\frac{1}{3}$    &    $\frac{2}{3}$    &    $\frac{1}{4}$    & 
0.56(4)    \\ 
O3    &    0.3311(15)    &    0.3039(19)    &    0.1722(4)    &    0.561(5) 
\\ 
D31    &    0.2226(18)    &    0.1022(19)    &    0.1774(7)    &    $=f(O3)$ 
\\ 
D32    &    0.2717(27)    &    0.3539(31)    &    0.1296(5)    &    $=f(O3)$ 
\\ 
O4    &    0.5714(21)    &    0.1036(26)    &    0.1849(6)    &    0.354(7) 
\\ 
D41    &    0.523(4)    &    -0.0911(28)    &    0.1823(12)    &    $=f(O4)$ 
\\ 
D42    &    0.6060(26)    &    0.178(5)    &    0.1378(8)    &    $=f(O4)$ 
\\ 
\hline 
\end{tabular} 
\begin{tabular}{llllll} 
Co1-O2 & 1.872(7) & Co2-O1 & 1.861(5) & Co2-O2 & 1.846(6) \\ 
Co2-O2 & 1.904 (8) & Na1-O3  & 2.352(8) & Na1-O4 & 3.030(4)  \\ 
Na2-O3  & 2.538(10) & Na2-O4  & 2.484(12) & D32-O2  & 1.680(11)\\ D42-O1 & 
1.945(16) & & & & \\ 
\hline 
\end{tabular} 
\end{center} 
\end{table} 
\endgroup 

\begin{figure}[!t] 
\includegraphics[natheight=750,natwidth=600,trim= 40 0 0 0,angle=90,scale=0.26]{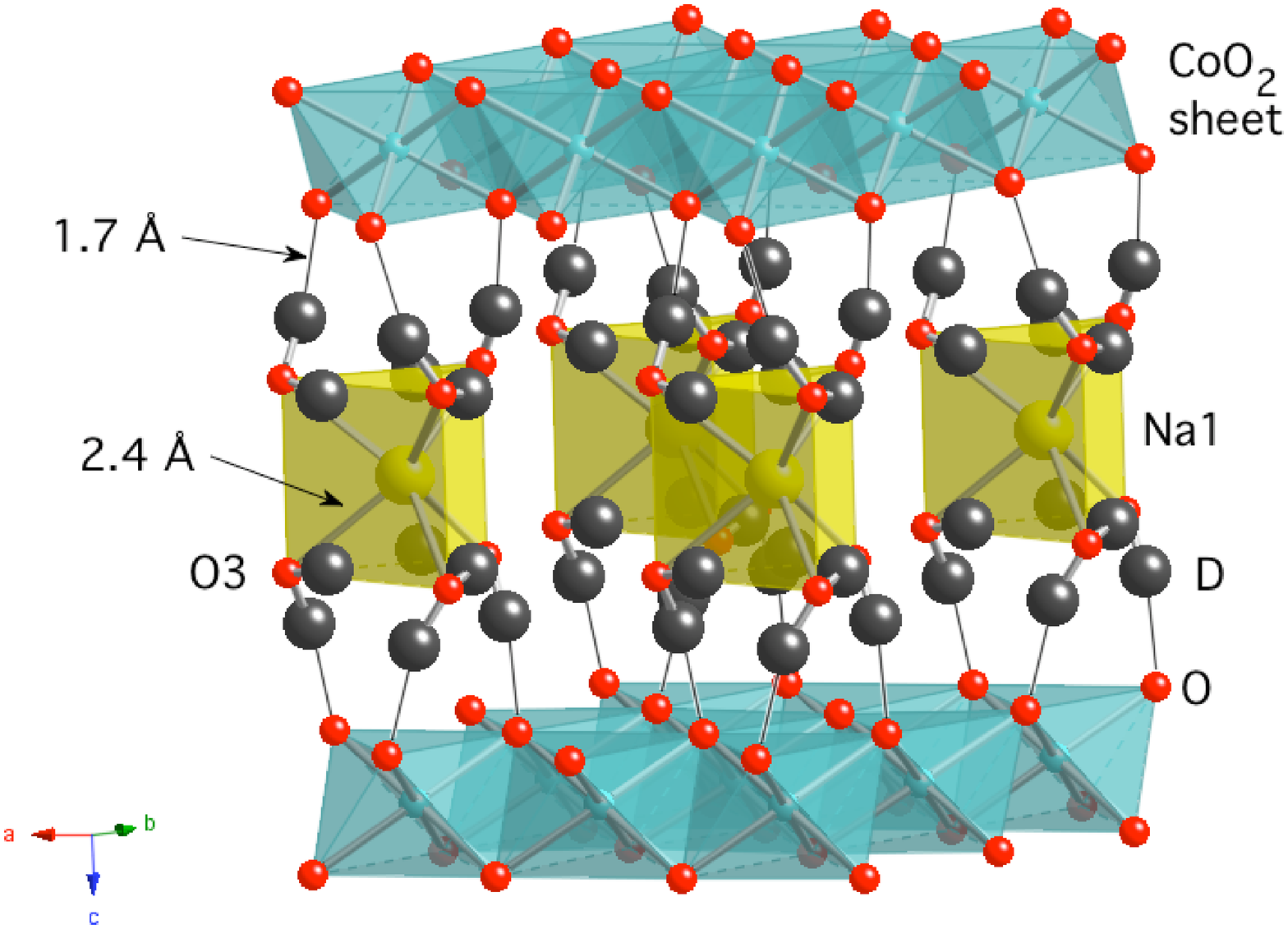} 
\caption{\footnotesize(Color) A three dimensional representation of the structure of 
\NaD\space obtained from our Rietveld analysis. For simplicity we show only half the unit cell along the $c$-axis and only the Na1 site with the \deut\space molecule centered at O3. The Na-O3 and D-O(-Co) bond-lengths are shown.} 
\label{strt1} 
\end{figure}

\begin{figure}[!t] 
\includegraphics[natheight=800,natwidth=550,trim=80 0 0 0,angle=-90,scale=0.35]{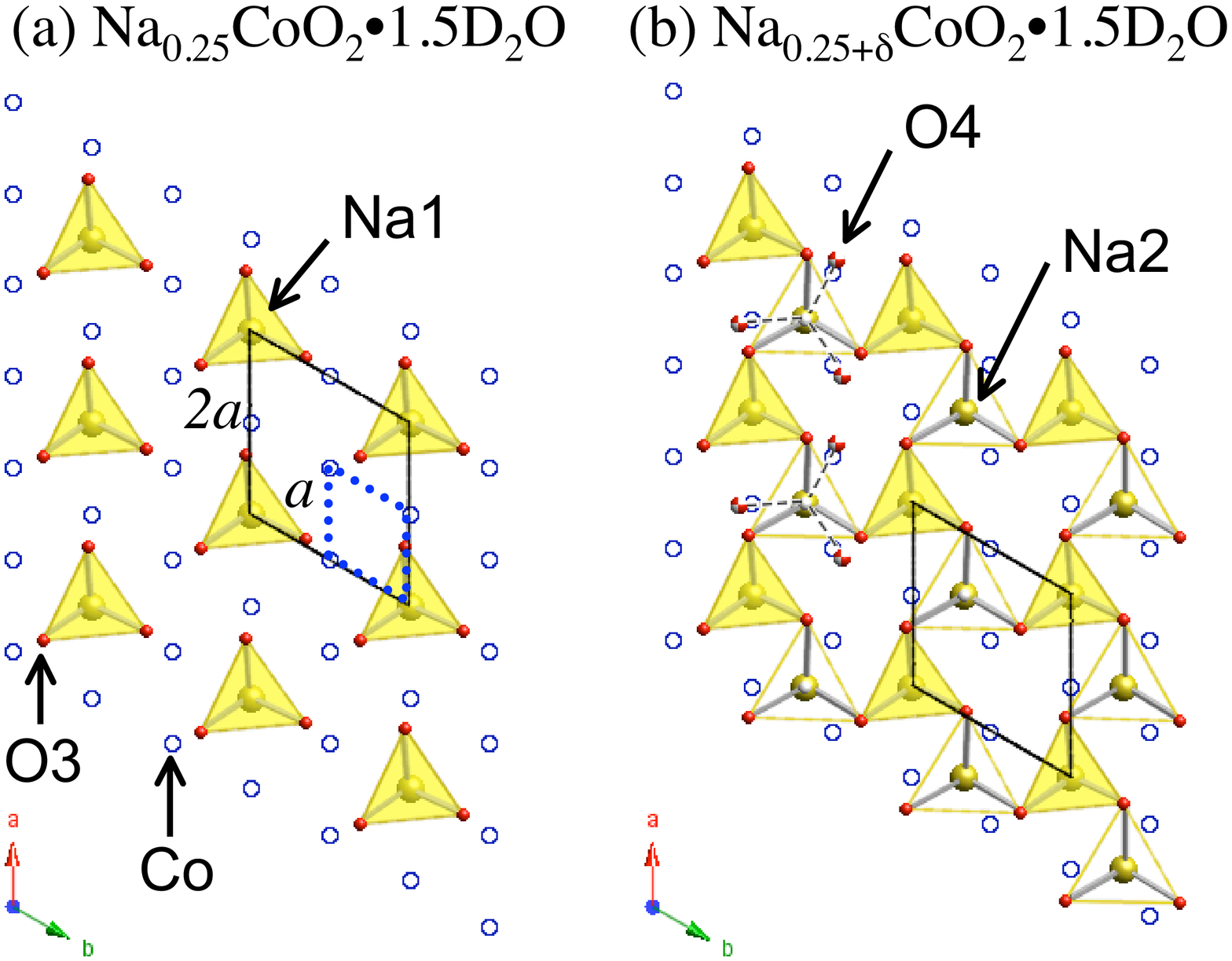} 

\caption{\footnotesize(Color) Projections of the Na-\deut\space layer down the $c-$axis. The NaO$_{6}$ triangular prisms centered on Na are highlighted. The \dhx\space unit cell is shown in solid lines and the $a \times a \times c$\space cell as dashed lines.  The doubling of the unit cell is evident from the position of the Na sites. D-atoms are omitted for clarity. (a) The idealized $x=\frac{1}{4}$ structure. (b) A model for the $x=\frac{1}{2}$ structure with the Na2 site filled. The coordination to the O4 site is shown. } 
\label{strt2} 
\end{figure}

The main structural features of our model of superconducting \NaD\space  are shown in figures ~\ref{strt1} and ~\ref{strt2}. The structure of CoO$_{2}$ sheets does not differ substantially from the one proposed by Takada.\cite{Takada}. To simplify the discussion of the Na and \deut\space environment, we first consider an ÒidealÓ structure of composition  $x$=$\frac{1}{4}$ and $y$=1.5 consisting of a fully occupied Na1 site surrounded by 6\deut\space molecules in a triangular prism geometry (see fig.~\ref{strt2}a). Each NaO$_{6}$ prism consists of a Na-atom in the center of the prism and O-atoms at each corner above and below the Na with Na-O bond-lengths of 2.4\AA\space (fig.~\ref{strt1}). For this composition, NaO$_{6}$ prisms would be uncoupled from each other. This is a very typical coordination for Na in Na-hydrate systems.\cite{Obst} In such materials Na atoms are surrounded by \wat\space forming \emph{hydration shells}, with the O-atom bonded to the 
alkali metal with a typical bond length of 2.4\AA, while the protons are 
pointing away from it.\cite{Obst} The most common Na-\wat\space coordination is 6, close to the value we find here. The doubling of $a$ and $b$ originates from this ordered arrangement of Na and its hydration shell (see fig.~\ref{strt2}). In fact, our NPD analysis has identified two Na sites centered at Na1 (0,0,$\frac{1}{4}$) and Na2 ($\frac{1}{3}$,$\frac{2}{3}$,$\frac{1}{4}$).   It is easy to see that additional Na (up to $x$=$\frac{1}{2}$) with the \emph{same} coordination number can be  accommodated into the ÒidealÓ structure by the progressive filling of the Na2 sites, without a substantial rearrangement of the water molecules. We find that the Na1 site is almost fully occupied with fractional  occupancy  $f\sim$ 0.88(4), while the second Na2 site is significantly less populated with  $f\sim$0.56(4). The coordination of 
O-atoms around the Na2 site is naturally the same as in the Na1, such that NaO$_{6}$ prisms can be centered at either Na site. The total amount of \deut\space determined from our Rietveld refinements is $y=$1.37(1), in good agreement with other reports.\cite{Takada,Foo,Jorgensen,Lynn}  Finally our analysis shows that at least one of the protons of the \deut\space molecule tend to point toward the CoO$_{2}$ sheets forming a hydrogen bond (D-O=1.7 \AA) with an O-atom in that layer (fig.~\ref{strt1}).  The structural model we employed in our best fits is slightly more complex than the one we just described, in that it includes a small amount of additional water centered around the O4 position (see table). This position, equivalent to the O3 position in the parent unit cell, may represent contributions of the four possible coherent domains of the double hexagonal cell. More likely, however, this site helps modeling the large amount of disorder in the structure (see below).

The main implication of our model, and the most relevant for the physics of this compound, is that \NaH\space has the \emph{potential} of being a homogeneous solid solution in a wide range of Na concentrations, rather than a point compound as suggested 
by Jorgensen \etal\cite{Jorgensen}.  A second important implication is that  changes of Na content can, within limits, be accommodated at constant water content, as each O-atom of the \deut\space molecule locally will be coordinated two Na-atoms.  In reality, a series of more or less stable ordered arrangements of Na2 occupied and vacant sites will occur for specific Na concentrations, although not all of them will be kinetically accessible.  The complex synthesis route of this material (de-intercalation of Na and 
intercalation of water) helps to explain why more than one superstructure type is observed in the same sample.  The inhomogeneous distribution of Na that is expected from 
de-intercalation produces a \emph{disordered background} on which 
intercalated \wat\space must arrange.   The intercalation of water would yield some re-organization of the Na-atoms at the local scale, but energetically a global re-arrangement is too costly (unless an anneal is possible).   As water is inserted into the lattice, local ordering of Na atoms occurs that resembles as much as possible the structure we describe. Fluctuations in the Na concentration are accommodated locally into the Na2-site, occasionally reaching an appropriate value for the formation of a long-period superstructure. This model would account for the predominant doubling mode and the many other diverse superstructures observed by ED. We note that our model does fold onto the structural models proposed thus far.\cite{Lynn, Jorgensen} For example the Na1 site of the model by Lynn \etal maps onto the Na1 site of our model and similarly for the model of Jorgensen \etal and the Na2 site.

The presence of these inhomogeneities should lead one to reconsider carefully the true nature of the superconducting phase. For example the predominant \dhx\space ordering may represent the 5K superconducting point phase while other supercells may also be superconducting with a lower \Tc. This scenario is  consistent with the almost constant variation of \Tc\space that has been observed as function $x$;\cite{Milne} NPD data over a range of $x$ that we have measured (0.28$<x<$0.37) display the 
($\frac{1}{2},\frac{1}{2},0$)-($\frac{1}{2},\frac{1}{2},1$) super-lattice 
reflection pair, indicating that the predominate of this ordering is quite 
stable over this $x$-range.  Other scenarios are possible, such as those in which a minority phase would be superconducting, and would partially shield the majority non-superconducting phase, but this is  unlikely as superconducting fractions are high, indicating that a majority phase is a superconductor. 

In summary, our neutron and electron diffraction data have unraveled a new aspect of the structure of \NaD\space, shedding   light into its material and physical properties. The formation of a NaO$_{6}$ triangular prism network can sustain a large range of $x$, without the necessity to vary the water content, or the separation between CoO$_{2}$ sheets. We argue on the basis of the structure that the amount of intercalated water essentially defines the separation between CoO$_{2}$ sheets and thus \Tc. A variety of sodium dopings, some of which leads to additional ordering can be accommodated in the structure and should have little influence on separation of CoO$_{2}$ sheets as Na can be distributed over two sites that coordinate to one water molecule, while maintain an average coordination close to the ideal Na-6\deut.

\end{document}